\documentclass{Interspeech2024}
\usepackage{xfrac}
\usepackage{cite}
\usepackage{nicefrac}
\usepackage{caption} 
\usepackage[table]{xcolor}
\usepackage{pifont}
\usepackage{multirow}
\usepackage{hhline}
\usepackage{xcolor}
\usepackage{tikz}

\usepackage{graphicx} 
\usepackage{dingbat} 
\usepackage{hyperref}

\newcommand{\inlinecircledigit}[1]{
  \tikz[baseline=(char.base)]{
    \node[shape=circle,draw,very thin,inner sep=1.2pt] (char) {#1};}
}

\interspeechcameraready 

\title{Reshape Dimensions Network for Speaker Recognition}

\name{Ivan}{Yakovlev}
\name{Rostislav}{Makarov}
\name{Andrei}{Balykin}
\name{Pavel}{Malov}
\name{Anton}{Okhotnikov}
\name{\ \ \ \ \ \ \ \ \ \ \ \ \ \ \ \ \ \ \ \ Nikita}{Torgashov}

\address{ID R\&D Inc., New York, USA}
\email{\tt\{yakovlev,makarov,andrew.balykin,pavel.malov,ohotnikov,torgashov\}@idrnd.net}


\begin{document}

\maketitle

\begin{abstract}
    

In this paper, we present \textbf{Re}shape \textbf{Dim}ensions \textbf{Net}work (ReDimNet), a novel neural network architecture for extracting utterance-level speaker representations. Our approach leverages dimensionality reshaping of 2D feature maps to 1D signal representation and vice versa, enabling the joint usage of 1D and 2D blocks. We propose an original network topology that preserves the volume of channel-timestep-frequency outputs of 1D and 2D blocks, facilitating efficient residual feature maps aggregation. Moreover, ReDimNet is efficiently scalable, and we introduce a range of model sizes, varying from 1 to 15 M parameters and from 0.5 to 20 GMACs. Our experimental results demonstrate that ReDimNet achieves state-of-the-art performance in speaker recognition while reducing computational complexity and the number of model parameters.


\end{abstract}

\section{Introduction}

Speaker recognition is a specialized field aiming at identifying or verifying individuals through their distinct voice features. In this domain, deep neural networks have emerged as a major technology for extracting speaker embeddings that are used for multiple tasks including Speaker Verification (SV), Speaker Identification, Speaker Diarization, and others. Extensive research has been conducted in the SV area, which includes the development of new datasets \cite{Vox2, VoxBlink, 3DSpeaker, VoxTube}, model architecture designing \cite{xvector, desplanques2020ecapa, next_tdnn, liu2022mfa, pcf_tdnn, dtdnn, Cam, zhang2022mfaconformer, thienpondt2024ecapa2, garcia2020magneto, resnext, eres2net, dfresnet, gemini}, and inventing new loss functions \cite{deng2019arcface, SF2}.



A variety of architectures have emerged including 1D \cite{xvector, desplanques2020ecapa, dtdnn, pcf_tdnn, next_tdnn} and 2D \cite{garcia2020magneto, resnext, eres2net, dfresnet, gemini} convolutional neural networks (CNNs), their hybrids that incorporate 2D CNN stem before 1D TDNN-like backbone \cite{liu2022mfa, Cam, thienpondt2024ecapa2}, as well as self-attention networks \cite{zhang2022mfaconformer}. Each architectural approach brings its unique set of advantages with 1D models offering efficiency and direct temporal analysis, 2D architectures providing frequency translational invariance \cite{thienpondt2021integrating}, and hybrid systems aiming to deliver the best of both worlds. Additionally, design approaches can be split into macro and micro designs, with micro designs involving modifications like substituting traditional 1D ResBlocks with Res2Net blocks within the ECAPA-TDNN architecture \cite{desplanques2020ecapa}, and macro designs incorporating a 2D stem ahead of TDNN-like models \cite{liu2022mfa, Cam, thienpondt2024ecapa2} leading to a two-stage architecture that transitions 2D $\xrightarrow{}$ 1D.



In this paper, we introduce ReDimNet\footnote{\url{https://github.com/IDRnD/ReDimNet}{}}, a novel neural network architecture based on the dimensionality reshaping of feature maps between 2D and 1D representations, enabling seamless integration of 1D and 2D blocks. ReDimNet exhibits scalability across various model sizes, while consistently achieving optimal performance under varying computational resource constraints. Our experimental results demonstrate that ReDimNet outperforms many other architectures and achieves state-of-the-art performance on public benchmarks while reducing inference time and model size.

\begin{figure}[!t]
  \centering
  \includegraphics[width=224pt]{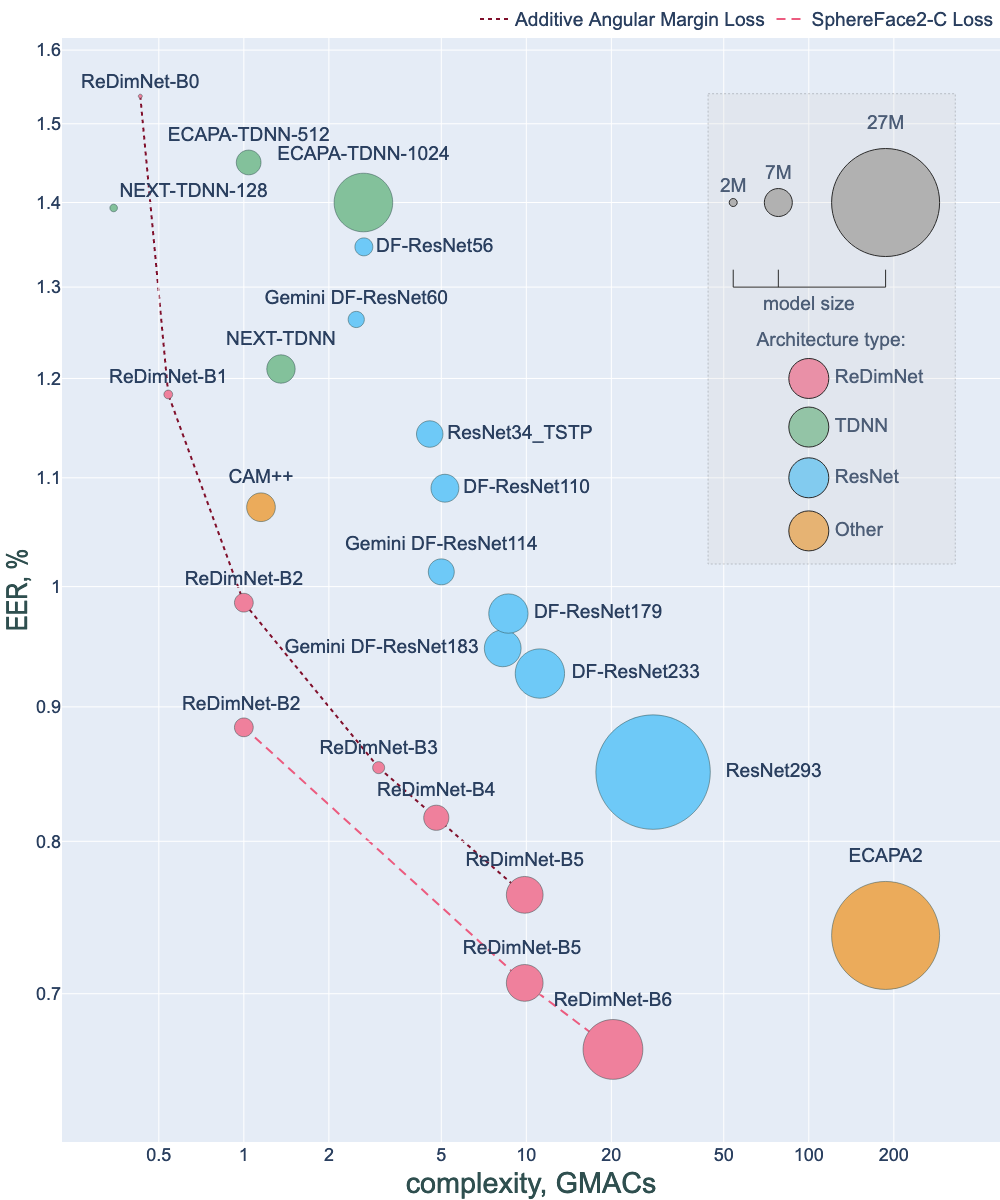}
  \caption{{\bfseries Computational Cost vs. Average Equal Error Rate.} EER is averaged over three Voxceleb1 protocols: Vox1-O, Vox1-E, Vox1-H. The model size is shown by the area of a circle, model family is indicated by a color. Complexity is assessed using \texttt{thop} library with an input signal of 2 seconds. A short dashed line represents scaling the ReDimNet architecture using the Additive Angular Margin loss function \cite{deng2019arcface}, dashed line - using the SphereFace2 loss \cite{SF2}.}
  \label{fig:speech_production}
\vspace{-10pt}
\end{figure}

\begin{figure*}[!t]
  \centering
  \includegraphics[width=\textwidth]{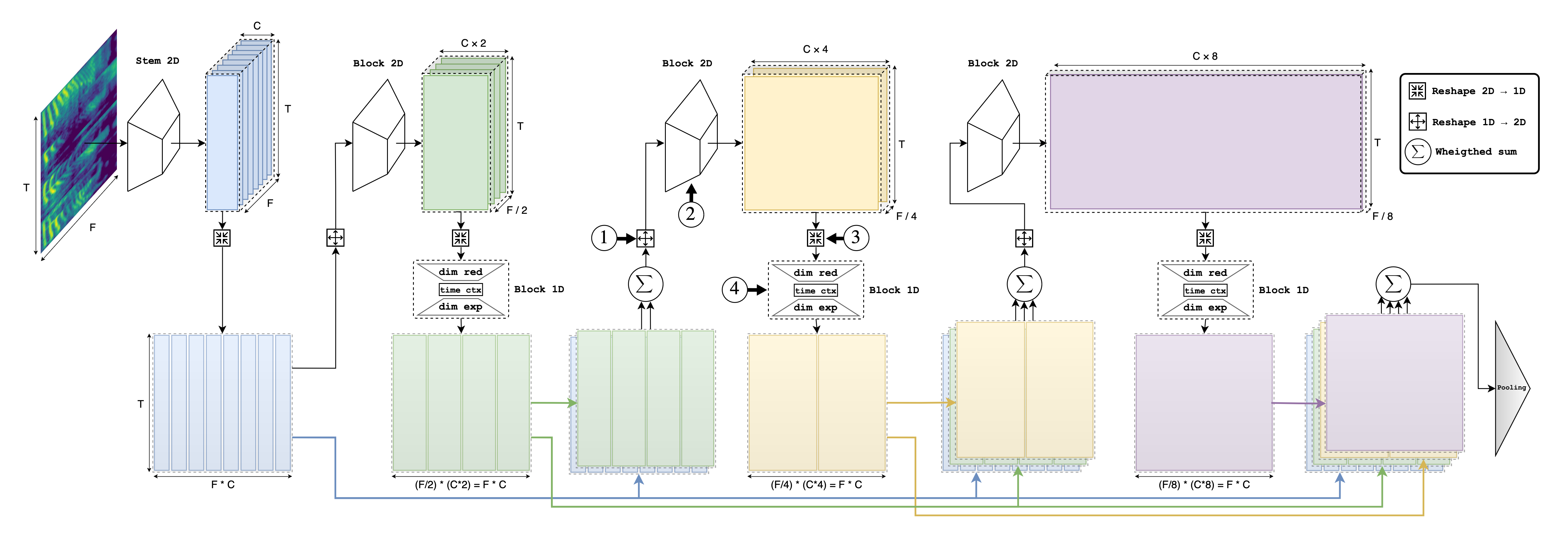}
  \caption{{\bfseries ReDimNet architecture scheme.} Digits 1,2,3 and 4 describe the order of operators and blocks execution in a single model stage, where $C$ - number of channels, $F$ - number of frequency bins, $T$ - number of timestamps.}
  \label{fig:model-scheme}
\end{figure*}



\section{Model Architecture}




In this section, we detail the design of the proposed architecture influenced by two main concepts. Firstly, to leverage the benefits of residual connections, we incorporate them extensively in ReDimNet. Secondly, based on the success of models utilizing both 1D and 2D blocks for speech processing and SV, our architecture integrates both types of blocks to boost performance.

\subsection{Dimensions reshape \& residual connection}

The main distinguishable feature of the architecture is an ability to aggregate 1D and 2D feature maps together with 2D feature maps from other model stages to enable 1D-2D and 2D-2D skip connections with various feature maps shapes. Such a technique is implemented only using the reshape operation without broadcasting or dimensionality reduction/expansion. We constrain ReDimNet to output feature maps with predefined shapes, that are easily reshaped back and forth between fixed 1D representation and various sets of 2D representations. First, we mitigate all strides across the time axis in a model, meaning that in the end, before pooling, the model will have the same time resolution as input features. Second, we synchronize strides over frequency dimension with a growth rate of channels for all stages, to make a "volume" of 2D feature maps constant throughout model forward pass. Given that, if all 2D feature maps are represented by the tensors with common PyTorch \cite{paszke2019pytorch} size notation $(s_{0} = batch\_size : bs, s_{1} = channels : C, s_{2} = frequncy : F, s_{3} = time : T)$, we assign volume of 2D feature map as $V = s_{1} \cdot s_{2} \cdot s_{3}$. This property of ReDimNet architecture is well illustrated in the model scheme (Fig. \ref{fig:model-scheme}) and in the Table \ref{feat-map-size-table} presenting the internal feature map size for each block.


\begingroup
\renewcommand{\arraystretch}{1.2} 
\begin{table}[htbp]
\centering
\caption{{\bfseries Model internal feature map sizes.} $S_{f}$ stands for frequency stride.}
\label{feat-map-size-table}
\scalebox{0.7}{
\begin{tabular}{c|c|c|c|c|c}
Block \# & In shape & $S_{f}$ & Channels & Out shape & Volume \\
\hline
1 & $(C, F, T)$ & 1 & $C$ & $(C, F, T)$ &  \\
2 & $(C, F, T)$ & 2 & $C \cdot 2$ & $(C \cdot 2, \nicefrac{F}{2}, T)$ &  \\
3 & $(C \cdot 2, \nicefrac{F}{2}, T)$ & 2 & $C \cdot 4$ & $(C \cdot 4, \nicefrac{F}{4}, T)$ & $C \cdot F \cdot T$ \\
4 & $(C \cdot 4, \nicefrac{F}{4}, T)$ & 2 & $C \cdot 8$ & $(C \cdot 8, \nicefrac{F}{8}, T)$ &  \\
5 & $(C \cdot 8, \nicefrac{F}{8}, T)$ & 1 & $C \cdot 8$ & $(C \cdot 8, \nicefrac{F}{8}, T)$ &  \\
\end{tabular}
}
\smallskip
\end{table}
\vspace{-10pt}
\endgroup

Having the same volume in all 2D feature maps is not yet enough to sum them right away to enable skip connections due to the shape mismatch. However, this can be easily overcome by an introduction of invertible reshape operator that reshapes all 2D feature maps of size $(bs, C_{i}, F_{i}, T)$ into 1D feature map of constant size: $(bs, C_{i} \cdot F_{i}, T) = (bs, C_{0} \cdot F_{0}, T)$, where $C_{0} = C$ and $F_{0} = F$. This equality is constant for various stages outputs due to the model strides and channels growth constraints. Then we sum 1D feature maps and reshape them back to 2D using the inverse reshape operator, this way we enable residual connection through the whole model forward pass.

\subsection{1D \& 2D Blocks}

ReDimNet is created around the use of joint 1D and 2D blocks, which are presented correspondingly by \texttt{Block 1D} and \texttt{Block 2D} in the scheme in Fig. \ref{fig:model-scheme}. These blocks are designed to handle 1D feature maps of fixed size, which are then reshaped into 2D feature maps for processing within the block. The structure of these blocks makes possible dynamic interchange between 2D and 1D representations: 2D subblocks process \inlinecircledigit{2} the reshaped (in \inlinecircledigit{1}) 1D inputs using sequences of residual blocks with 2D convolutions, and then the output is converted back to a 1D format \inlinecircledigit{3} for further processing in the 1D subblock \inlinecircledigit{4}. This 1D subblock employs a channel-axis dimensionality reduction Fully Connected (FC) layer + normalization layer, followed by a time-contextual processing component. This component can be implemented through ConvNeXt-like 1D blocks, transformer encoder blocks, or a combination of them, and its output is a 1D feature map. Finally, the channel-axis expansion FC layer unfolds the number of channels to match input shape and performs skip+residual sum operation. More information on the basic blocks structure used in ReDimNet is provided in Fig. \ref{fig:micro_block_design}.

\subsection{Input features \& pooling}

As model input features we used 72-dimensional mean-normalized Mel filter bank log-energies with a 25 ms frame length and 15 ms step with 512 FFT size over the 20-7600 Hz frequency range by default. To extract an utterance-level embedding from the frame-level features, we utilized the Attentive Statistics Pooling \cite{att_pool} with global context.


\section{Experimental Setup}

We conducted experiments of training ReDimNet architecture utilizing the development part of the VoxCeleb2\cite{Vox2} dataset. Models were optimized using SGD optimizer with Nesterov momentum, $m = 0.9$, and a weight decay of $2e^{-5}$. As a default loss function, we selected Additive Angular Margin (AAM) softmax loss \cite{deng2019arcface} due to its wide adoption. We also conducted and reported results for a few experiments with SphereFace2 (SF2) loss function \cite{SF2} for comparison purposes. We followed a 2-stage training approach by firstly pretraining a model on short segments with multiple augmentations applied. Then, we applied finetuning on longer utterances with some augmentations turned off and tweaked the parameters of a loss function. This second training stage is well-known as Large-Margin (LM) finentuning strategy \cite{LM}. All models were trained using the \texttt{wespeaker} \cite{wang2023wespeaker} training pipeline.

\subsection{Pretraining stage}

For pretraining, we used a default \textit{voxceleb2} recipe from \texttt{wespeaker} pipeline with minor adjustments. 2-second segments were selected randomly from each signal, and various augmentations with MUSAN dataset \cite{snyder2015musan} (noise, music, babble) alongside the RIR dataset \cite{szoke2019building} were applied following the augmentation recipe from \cite{garcia2020magneto}. A two-fold speed augmentation \cite{ko2015audio}, with factors of 0.9 and 1.1, was employed to generate additional speakers within the training dataset. In this stage, the AAM-softmax margin penalty was scheduled as follows: first 20 epochs it was kept at $0.0$, then for the next 20 epochs it exponentially rose to 0.2 and then was kept constant till the end of training. We used Exponential Decay with Warmup learning rate scheduler with 6 epochs warmup, $lr_{max}=1e^{-1}$ and $lr_{min}=1e^{-5}$.

\subsection{Large-Margin Finetuning stage}
At the finetuning stage \cite{LM}, AAM-softmax margin was set to constant 0.5 value, with length of training utterances expanded to 6 seconds. Speed perturbations were turned off during this stage.

\begin{figure}[!t]
  \centering
  \includegraphics[width=230pt]{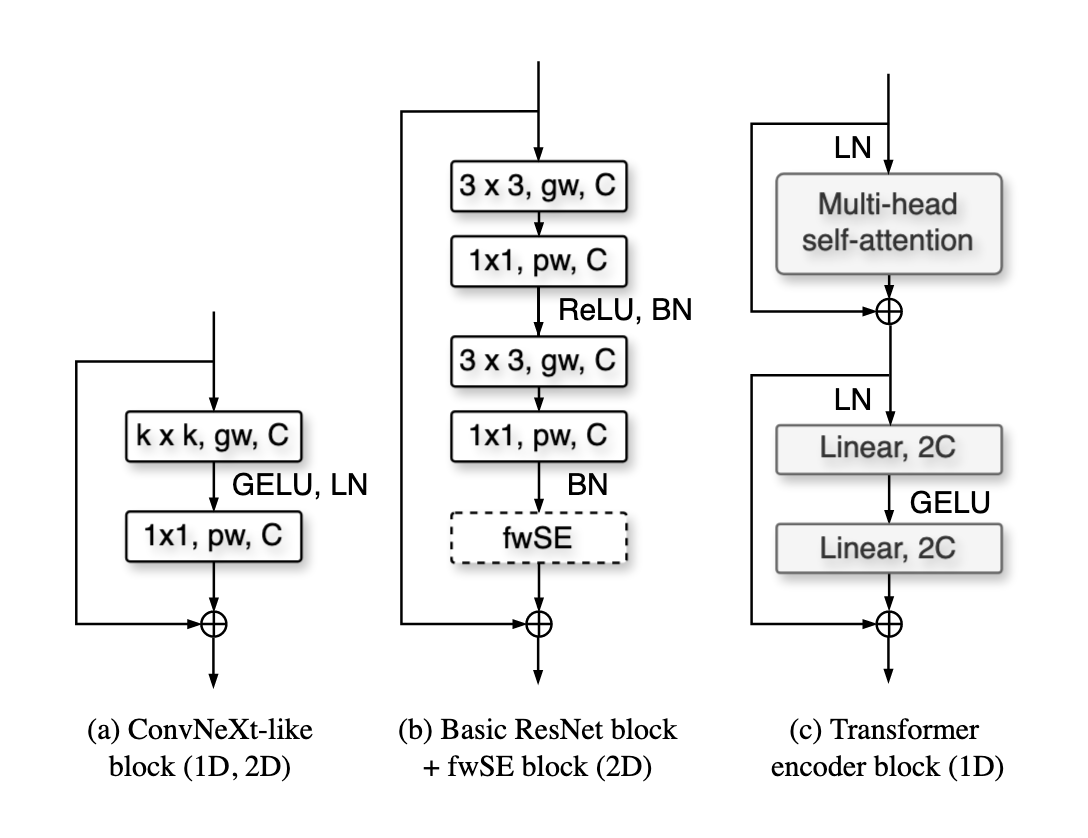}
  \caption{{\bfseries Block design.} In ReDimNet as 2D blocks we used (a) slightly modified ConvNeXt-like block \cite{liu2022convnet} or (b) basic ResNet block \cite{resnet} with fwSE \cite{thienpondt2021integrating}. As 1D blocks we used same (a) 1D version of ConvNeXt-like block with or inplace of (c) Transformer block \cite{vaswani2023attention}.}
  \label{fig:micro_block_design}
  \vspace{-10pt}
\end{figure}





\subsection{Evaluation}
The performance of models is assessed using cleaned protocols of VoxCeleb1 \cite{Nagrani19} test set, employing the Equal Error Rate (EER) and the minimum Detection Cost Function (minDCF) with $P_{target} = 0.01$ and $C_{FA} = C_{Miss} = 1$. We scored each model with cosine backend utilizing full utterance length as input and additionally applied a top-300 adaptive s-normalization (AS-Norm) \cite{asnorm} of cosine scores (see Table \ref{tab:resnet-comparison}).



\section{Model scaling \& ablation studies}

\subsection{Model scaling}
Achieving efficient model scaling was one of our main research goals. Therefore, we were able to scale the ReDimNet architecture from (1M, ~0.5 GMACs) to (15M, 20 GMACs), reaching competitive results for each model size. For the naming convention, we followed notations of \cite{tan2020efficientnet}, resulting in 7 configurations: B0 - B6, where each model configuration is bounded by the computational complexity limits in GMACs that we found to be a predominant factor of model scaling relative to model size. Complete testing results of each ReDimNet configuration are shown in Table \ref{tab:resnet-comparison}.

\begingroup
\begin{table}[ht]
\centering
\caption{Ablation Study on Block Components of ReDimNet (EER, \%)}
\label{table:ablation_study_refined}
\resizebox{\columnwidth}{!}{%
\begin{tabular}{clcccc}
\hline
& \textbf{Block Type} & \textbf{Vox1-O} & \textbf{Vox1-E} & \textbf{Vox1-H} & \textbf{Average} \\ \hline
\multirow{5}{*}{\rotatebox[origin=c]{90}{\scriptsize\textit{1D block}}} & Skip Connection  & 1.59          & 1.57           & 2.71           & 1.96           \\
& Fully Connected   & 0.93           & 1.13           & 1.94           & 1.33            \\
& 1D Conv    & 0.65           & 0.85           & 1.54            & 1.01            \\
& MHA      & 0.69           & 0.82           & \textbf{1.45}           & 0.99            \\
& 1D Conv + MHA  & \textbf{0.59}         & \textbf{0.79}           & 1.47           & \textbf{0.95}            \\
\hline
\multirow{3}{*}{\rotatebox[origin=c]{90}{\scriptsize\textit{2D block}}} 
& ConvNext block              & 0.68            & 0.83            & \textbf{1.46}            & 0.99            \\ 
& fwse-ResNet block               & 0.64            & 0.82            & 1.48            & 0.98            \\
& ResNet block                    & \textbf{0.61}            & \textbf{0.80}             & 1.48            & \textbf{0.96}           \\
\hline
\end{tabular}%
}
\end{table}
\vspace{-10pt}
\endgroup

\subsection{Ablation studies}

We also conducted a thorough study of how different components of ReDimNet architecture affect its performance. This research includes studying the role of 1D and 2D blocks for speech signal processing, assessing the impact of different loss functions, and optimizing group sizes and steps in convolutions for accuracy  and efficiency improvement. All ablation studies were carried out on the ReDimNet-B2 architecture.

\begin{table}[ht]
\centering
\caption{Ablation study on loss function configuration (EER,\%)}
\label{table:loss_ablation}
\resizebox{\columnwidth}{!}{%
\begin{tabular}{lcccc}
\hline
\textbf{Loss Type} & \textbf{Vox1-O} & \textbf{Vox1-E} & \textbf{Vox1-H} & \textbf{Average} \\ \hline
AAM-SC           & \textbf{0.57}          & 0.91           & 1.60           & 1.03         \\ 
AAM                & 0.68           & 0.83           & 1.46           & 0.99            \\
SF2-A              & 0.63          & 0.80           & 1.39           & 0.94            \\
SF2-C              & \textbf{0.57}           & \textbf{0.76}           & \textbf{1.32}           & \textbf{0.88}            \\
\hline
\end{tabular}%
}
\end{table}
\vspace{-10pt}

\begingroup
\newcommand{\sr}{\rule[-0.25cm]{0pt}{0.7cm}}
\begin{table*} 
\centering
\caption{{\bfseries Evaluation results on the VoxCeleb1-Cleaned protocols without QMFs.} For the report, we calculated the equal error rate (EER) and the minimum detection cost function (minDCF). GMACs were measured on 2-s long segments. * - means values have been estimated. Open source models from the WeSpeaker or ECAPA2 repositories were retested in our environment.}
\label{tab:resnet-comparison}
\setlength\tabcolsep{4pt} 
\renewcommand{\arraystretch}{0.9} 
\resizebox{0.9\textwidth}{!}{
\begin{tabular}{lllcccccccc} 
\hline\hline
  \multirow{2}{*}{\textbf{Model}} &
  \multirow{2}{*}{\textbf{Params}} &
  \multirow{2}{*}{\textbf{GMACs}} &
  \multirow{2}{*}{\textbf{LM}} &
  \multirow{2}{*}{\textbf{AS-Norm}} &
  \multicolumn{2}{c}{{\textbf{Vox1-O}}} &
  \multicolumn{2}{c}{{\textbf{Vox1-E}}} &
  \multicolumn{2}{c}{{\textbf{Vox1-H}}} \\ 
  \textbf{} &
  \textbf{} &
  \textbf{} &
  \textbf{} &
  \textbf{} &
  \textbf{EER(\%)} &
  \textbf{minDCF} &
  \textbf{EER(\%)} &
  \textbf{minDCF} &
  \textbf{EER(\%)} &
  \textbf{minDCF} \\ \hline

\textbf{ReDimNet-B0}                            & \multirow{2}{*}{\textbf{\hspace{1.5mm}1.0M}} & \multirow{2}{*}{\textbf{\hspace{3mm}0.43}} &\ding{51}&\ding{55} &1.16          &0.101          &1.25          &0.132          &2.20 &0.207    \\
%
\hspace{0.5cm}+AS-Norm  & & &\ding{51}&\ding{51} &\textbf{1.07} &\textbf{0.098} &\textbf{1.18} &\textbf{0.121}    &\textbf{2.01}      &\textbf{0.184}    \\
\hline 
NeXt-TDNN-l (C=128, B=3)\cite{next_tdnn}      &\textbf{\hspace{1.5mm}1.6M}      &\textbf{\hspace{3mm}0.29}{*}     &\ding{55}&\ding{51}&1.10      &0.108    &1.24      &0.133    &2.12      &0.201    \\
NeXt-TDNN (C=128, B=3)\cite{next_tdnn}        &\hspace{1.5mm}1.9M      &\hspace{3mm}0.35{*}     &\ding{55}&\ding{51}&1.03      &0.095    &1.17      &0.126    &1.98      &0.190    \\
\textbf{ReDimNet-B1 }                  & \multirow{2}{*}{\hspace{1.5mm}2.2M}      & \multirow{2}{*}{\hspace{3mm}0.54}      &\ding{51}&\ding{55}&0.85      &0.076    &0.97      &0.104    &1.73      &0.166    \\
\hspace{0.75cm}+AS-Norm       &          &          &\ding{51}&\ding{51}&    \textbf{0.73}      &    \textbf{0.071}      &     \textbf{0.89}     &      \textbf{0.096}    &     \textbf{1.57}     &     \textbf{0.154}     \\
\hline 
 ECAPA (C=512)\cite{desplanques2020ecapa, dfresnet}                 &\hspace{1.5mm}6.4M      &\hspace{3mm}1.05      &\ding{55}&\ding{51}&0.94      &0.092    &1.21      &0.129    &2.20      &0.205    \\
 NeXt-TDNN-l (C=256, B=3)\cite{next_tdnn}      &\hspace{1.5mm}6.0M      &\hspace{3mm}1.13{*}     &\ding{55}&\ding{51}&0.81      &0.091    &1.04      &0.116    &1.86      &0.184    \\
 CAM++\cite{Cam, wang2023wespeaker}                &\hspace{1.5mm}7.2M      &\hspace{3mm}1.15      &\ding{51}&\ding{55}&0.71      &0.109    &0.85      &0.095    &1.66      &0.165    \\
 NeXt-TDNN (C=256, B=3)\cite{next_tdnn}        &\hspace{1.5mm}7.1M      &\hspace{3mm}1.35{*}     &\ding{55}&\ding{51}&0.79      &0.087    &1.04      &0.115    &1.82      &0.182    \\
 $\textbf{ReDimNet-B2}_{SF2}$            & \multirow{2}{*}{\textbf{\hspace{1.5mm}4.7M}}      & \multirow{2}{*}{\textbf{\hspace{3mm}0.90}}      &\ding{51}&\ding{55}&0.57      &0.054    &0.76      &0.082    &1.32      &0.133    \\
\hspace{0.75cm}+AS-Norm     &          &      &\ding{51}&\ding{51}&         \textbf{0.52} &   \textbf{0.060}       &   \textbf{ 0.74  }    &   \textbf{ 0.078  }    &   \textbf{ 1.27 }     &   \textbf{ 0.128 }     \\
\hline 
 ECAPA (C=1024)\cite{desplanques2020ecapa, dfresnet}                &14.9M     &\hspace{3mm}2.67      &\ding{51}&\ding{55}&0.98      &0.105    &1.13      &0.117    &2.09      &0.204    \\
 DF-ResNet56\cite{dfresnet}                &\hspace{1.5mm}4.5M     &\hspace{3mm}2.66      &\ding{55}&\ding{51}&0.96      &0.103    &1.09      &0.122    &1.99      &0.184    \\
 Gemini DF-ResNet60\cite{gemini}                 &\hspace{1.5mm}4.1M     &\textbf{\hspace{3mm}2.50}{*}      &\ding{55}&\ding{51}&0.94      &0.089    &1.05      &0.116    &1.80      & 0.166    \\

  \textbf{ReDimNet-B3}                   & \multirow{2}{*}{\textbf{\hspace{1.5mm}3.0M}}      & \multirow{2}{*}{\hspace{3mm}3.00}      &\ding{51}&\ding{55}&0.50      &0.063    &0.73      &0.079    &1.33      &0.135    \\
  \hspace{0.75cm}+AS-Norm        &          &          &\ding{51}&\ding{51}&\textbf{0.47}      &\textbf{0.042}    &\textbf{0.69}      &\textbf{0.072}    &\textbf{1.23}      &\textbf{0.121}    \\
\hline 
 ResNet34\cite{wang2023wespeaker, resnet}                      &\hspace{1.5mm}6.6M     &\textbf{\hspace{3mm}4.55}      &\ding{51}&\ding{55}&0.82      &0.080    &0.93      &0.104    &1.68      &0.161    \\
 Gemini DF-ResNet114\cite{gemini}           &\hspace{1.5mm}6.5M      &\hspace{3mm}5.00      &\ding{55}&\ding{51}&0.69      &0.067    &0.86      &0.097    &1.49      &0.144    \\
 \textbf{ReDimNet-B4}                   & \multirow{2}{*}{\textbf{\hspace{1.5mm}6.3M}}      & \multirow{2}{*}{\hspace{3mm}4.80}      &\ding{51}&\ding{55}&0.51      &0.052    &0.68      &0.073    &1.26      &0.123    \\
  \hspace{0.75cm}+AS-Norm        &          &          &\ding{51}&\ding{51}&\textbf{0.44}      &\textbf{0.042}    &\textbf{0.64}      &\textbf{0.067}    &\textbf{1.17}      &\textbf{0.111}    \\
\hline 
 Gemini DF-ResNet183\cite{gemini}           &\hspace{1.5mm}9.2M      &\textbf{\hspace{3mm}8.25}      &\ding{55}&\ding{51}&0.60      &0.064    &0.81      &0.090    &1.44      &0.137    \\
 DF-ResNet233\cite{dfresnet}                 &12.3M     &\hspace{1.5mm}11.17     &\ding{55}&\ding{51}&0.58      &0.044    &0.76      &0.083    &1.44      &0.146    \\
 $\textbf{ReDimNet-B5}_{SF2}$             & \multirow{2}{*}{\textbf{\hspace{1.5mm}9.2M}}      & \multirow{2}{*}{\hspace{3mm}9.87}      &\ding{51}&\ding{55}&0.43      &0.039    &0.61      &0.062    &1.08      &0.102    \\
  \hspace{0.75cm}+AS-Norm        &          &          &\ding{51}&\ding{51}&\textbf{0.39}      &\textbf{0.037}    &\textbf{0.59}      &\textbf{0.057}    &\textbf{1.05}      &\textbf{0.095}    \\
\hline 
 ResNet293\cite{wang2023wespeaker, resnet}                     &23.8M    &\hspace{1.5mm}28.10      &\ding{51}&\ding{55}&0.53      &0.057    &0.71      &0.072    &1.30      &0.127    \\
 ECAPA2\cite{thienpondt2024ecapa2}                       &27.1M     & 187.00{*}    &\ding{51}&\ding{55}&0.44      &0.041    &0.62      &0.066    &1.15      &0.114    \\
 $\textbf{ReDimNet-B6}_{SF2}$                    & \multirow{2}{*}{\textbf{15.0M}}    & \multirow{2}{*}{\textbf{\hspace{1.5mm}20.27}}     &\ding{51}&\ding{55}&0.40      &0.033    &0.55      &0.052    &1.05      &0.104    \\
  \hspace{0.75cm}+AS-Norm               &          &          &\ding{51}&\ding{51}&\textbf{0.37}      &\textbf{0.030}    &\textbf{0.53}      &\textbf{0.051}    &\textbf{1.00}      &\textbf{0.097}    \\
\hline\hline
\end{tabular}
}
\vspace{0pt}
\end{table*}
\endgroup

\subsubsection{Block types}
In order to identify optimal configurations of ReDimNet architecture, we compared three types of 2D-blocks: basic ResNet block, basic ResNet FWSE block, and a ConvNext block. While minimal differences were observed, basic ResNet block, however, slightly outperformed others by a small margin (see Table \ref{table:ablation_study_refined}).

Our further analysis was focused on the 1D block type, where we assessed a range of options including sequences of 1D convolutional ConvNeXt-like blocks (Fig. \ref{fig:micro_block_design}), multi-head attention (MHA) (Fig. \ref{fig:micro_block_design}), FC layers, skip connections, and a hybrid of 1D convolutional blocks with MHA (1D Conv + MHA). Skip connections appeared to be the least effective approach, which underscored the importance of the 1D block within the ReDimNet architecture. FC layers performed slightly better, suggesting the importance of a temporal context. 1D convolutional and MHA blocks have proven to be the most efficient configurations, and a combination of MHA and 1D convolutional blocks delivered the best performance (see Table \ref{table:ablation_study_refined}).


\subsubsection{Loss studies}

Furthermore, we explored the effectiveness of various loss functions (see Table \ref{table:loss_ablation}). Specifically, we evaluated SphereFace losses (SF2) with A and C configurations \cite{SF2}, Additive Angular Margin Loss (AAM), and Additive Angular Margin loss with SubCenters (AAM-SC) \cite{deng2019arcface}. Based on the testing results, we found SphereFace type C to be the most effective loss function providing the largest performance improvement in the benchmarks.

\section{Results}

Testing results of all proposed ReDimNet architecture configurations are presented in Table \ref{tab:resnet-comparison}. We compared ReDimNet on the VoxCeleb1 protocols with publicly available models and grouped them based on the number of parameters and multiply-accumulate operations (MACs) for comparison purposes. 

In particular, our ReDimNet-B1 model achieves comparable results to NeXt-TDNN \cite{next_tdnn} on the Vox1-H protocol, but has a slightly larger number of parameters and MACs. ReDimNet-B3 outperforms Gemini DF-ResNet60 \cite{gemini} and ECAPA (C = 1024) with an advantage in model size. ReDimNet-B5 further improves upon the B3 version, consistently achieving the lowest EER and minDCF, compared to DF-ResNet233 \cite{dfresnet}, which has the similar number of parameters and MACs. Moreover, our largest model, ReDimNet-B6, delivers even better results while having significantly fewer parameters and MACs than ECAPA 2 \cite{thienpondt2024ecapa2} and ResNet293 \cite{wang2023wespeaker, resnet}.

Furthermore, we subjected the best models of various architectures to additional out-of-domain testing (see Table \ref{table:other_tests}). These results demonstrate that ReDimNet-B6 outperforms other architectures with a significant gap on unseen data domains.


\begin{table}[ht]
\centering
\caption{Evaluation results on Speakers In The Wild core-core protocol \cite{mclaren2016speakers}, VOiCES from a Distance Challenge Evaluation Set \cite{nandwana2019voices} and VoxCeleb1-B protocol \cite{nam2023disentangled} (EER, \%).}
\label{table:other_tests} 
\setlength{\tabcolsep}{4pt} 
\begin{tabular}{lcccc}
\hline
\textbf{Model} & \textbf{SITW} & \textbf{VOiCES} & \textbf{Vox1-B} & \textbf{Average} \\ \hline
CAM++ & 1.34 & 6.30 & 2.79 & 3.48  \\
ECAPA (C=1024) & 1.67 & 5.31 & 3.48 & 3.49 \\
ResNet293 & 1.67 & 5.14 & 2.23 & 3.01 \\
ECAPA2 & 3.64 & 13.26 & 1.81 & 6.24 \\
\textbf{ReDimNet-B6} & \textbf{0.77} & \textbf{3.19} & \textbf{1.66} & \textbf{1.87} \\ \hline
\end{tabular}
\label{tab:adjusted_highlighted}
\end{table}
\vspace{-10pt}

\section{Conclusions}

In this paper we introduced ReDimNet - a novel neural network architecture designed for the extraction of utterance-level speaker representations. It combines dimensionality reshaping, dynamic transitions between 1D and 2D representations, and 2D and 1D blocks.
Through a comprehensive evaluation, \ \ \ ReDimNet demonstrated:
\begin{itemize}[leftmargin=15pt]
    \item architecture adaptability and scalability across multiple configurations;
    \item top balance between computational efficiency and performance;
    \item strong results on the VoxCeleb1-H (cleaned) protocol, with an Equal Error Rate (EER) of 1.00\%;
    \item advanced generalization ability on out-of-domain test sets.
\end{itemize}

In summary, ReDimNet architecture achieves competitive performance on all tests compared to other state-of-the-art speaker recognition models, while also offering favorable computational efficiency. Its adaptability and superior performance make it a valuable contribution to the speaker recognition field and a promising solution for real-world applications.

\newpage
\bibliographystyle{IEEEtran}
\bibliography{mybib}

\end{document}